\documentclass[10pt, conference, letterpaper]{IEEEtran}

\ifCLASSINFOpdf
\else
\fi
\usepackage{url}
\usepackage{amsmath, amsfonts,  amssymb, color, textcomp}
\usepackage[all]{xy}
\usepackage{graphicx}
\usepackage{latexsym}
\usepackage{epsfig}
\usepackage{bm}
\usepackage[footnotesize]{caption}
\usepackage{float}
\usepackage{blindtext}
\usepackage{tikz}
\usetikzlibrary{fit,positioning}
\usetikzlibrary{shapes,matrix,decorations.markings,arrows}
\usetikzlibrary{graphs}
\usepackage{setspace}

\IEEEoverridecommandlockouts

\definecolor{red}{rgb}{1,0,0}
\usepackage{color,soul}

\definecolor{lightblue}{rgb}{.90,.95,1}
\sethlcolor{yellow}

\usepackage{psfrag}

\long\def\symbolfootnote[#1]#2{\begingroup%
\def\thefootnote{\fnsymbol{footnote}}\footnote[#1]{#2}\endgroup} 

\usepackage[outdir=./]{epstopdf}

\begin{document}

%Start of paper content:

\title{Iterative Threshold Decoding of Spatially Coupled, Parallel-Concatenated Codes\vspace{-3mm}}

\author{
\IEEEauthorblockN{Andrew D. Cummins$^\dag$, David~G.~M.~Mitchell$^\dag$, and Daniel~J.~Costello,~Jr.$^\ddag$}\vspace{2mm}
\IEEEauthorblockA{$^\dag$Klipsch School of Electrical and Computer Engineering, New Mexico State University, Las Cruces, NM\\
$^\ddag$Dept. of Electrical Engineering, University of Notre Dame, Notre Dame, IN\\\{andrewdc, dgmm\}@nmsu.edu, costello.2@nd.edu\\}
\vspace{-9mm}}

\maketitle

\begin{abstract}
\normalsize
Spatially coupled, parallel concatenated codes (SC-PCCs) have been shown to approach channel capacity when decoded using optimal iterative methods.  However, under complexity constraints such decoding strategies can result in unacceptable power and latency costs.  In this work, we employ convolutional self-orthogonal component codes along with low-complexity, suboptimal a posteriori probability (APP) threshold decoders with SC-PCCs to reduce decoding complexity.  The proposed code design is faster, more energy efficient, and easier to implement than optimal methods, while offering significant coding gain over existing threshold decodable, turbo-like constructions of similar latency and complexity.  The design also serves to further illustrate the advantages spatial coupling can provide to existing code constructions and decoder implementations. 
\end{abstract}\vspace{-5mm}

%%%%%%%%%%%%%%%%%%%%%%%%%%%%%%%%%%%%%%%%%%%%

\section{Introduction}\label{sec:intro}
Coding\symbolfootnote[0]{This material is based upon work supported by the National Science Foundation under Grant Nos. OIA-1757207 and HRD-1914635.} schemes for high-throughput communication systems often rely on hard-decision (\emph{i.e.}, syndrome-based) component decoder implementations and are typically unable to take advantage of soft reliability information during decoding, due to the requisite order-of-magnitude increase in computational overhead and/or power consumption \cite{zhangLowComplexitySoftDecisionConcatenated2017}. However, in \cite{riedelIterativeTurboDecoding1995} the authors proposed that \emph{parallel concatenated convolutional} (PCC) or \textit{turbo} codes can be decoded iteratively with low complexity, soft-decision \emph{a posteriori probability} (APP) threshold decoders if the component codes are \emph{convolutional self-orthogonal codes} (CSOCs). The growth in decoding complexity of such schemes is linear with respect to the constraint length of the component codes, making them attractive for high-throughput, low-overhead applications \cite{fangMethodDeviceError2002}. Additional schemes for efficient iterative decoding of CSOCs with similar trade-offs were proposed in \cite{heProceduresEfficientIterative2006}.

Codes on graphs whose constraint nodes are convolutional codes are referred to as \emph{spatially coupled, turbo-like codes} (SC-TCs) \cite{hzc08,moloudiSpatiallyCoupledTurboLike2017,moloudiSpatiallyCoupledTurbolike2019,yangPartiallyInformationCoupledTurbo2018}. These codes have been shown to exhibit the threshold saturation phenomenon, wherein their performance under iterative belief propagation (BP) decoding approaches the decoding threshold of  optimal \emph{maximum a posteriori} (MAP) decoding of the uncoupled code ensemble \cite{moloudiSpatiallyCoupledTurboLike2017}. Such decoding schemes typically employ BCJR component decoders, the complexity of which could be a limiting factor for resource constrained implementations. 

The motivation of this paper is to investigate if low complexity, high speed schemes involving spatial coupling can close the performance gap between optimal and suboptimal decoded PCCs. We therefore consider a low-overhead SC-TC with CSOC component codes whose structure is a direct extension of the familiar PCC, and we provide a low-complexity soft-decision APP threshold decoding algorithm for these \emph{spatially coupled} PCCs (SC-PCCs). This scheme enables the use of high rate and large constraint length component codes with acceptable complexity. The encoder for a CSOC SC-PCC is identical to that of a PCC, with additional coupling of source information blocks prior to encoding. The proposed SC-PCC decoder employs a sliding-window type procedure in which each block is decoded by two component APP threshold decoders similar to those described for PCCs \cite{riedelIterativeTurboDecoding1995}, but spread over time. We demonstrate significant coding gain over non-spatially coupled APP threshold-decodable PCC codes with  comparable latency and decoding complexity.

\section{CSOCs and Threshold Decoding}\label{sec:secSCPCC_01}

A $(k+1, k, m)$ convolutional code in systematic form is said to be \emph{self-orthogonal} with error-correcting capability $\lfloor \frac{J}{2} \rfloor$ if, for each information error symbol $e_l^{(i)}$ at time $l$, $i\in\{0,1,\ldots,k-1\}$, the $J$ parity-check equations $A_j^{(i)}$, $j\in\{1,2,\ldots,J\}$, that check $e_l^{(i)}$ do not check any other error symbol more than once. 
At the decoder, sets of $J$ self-orthogonal checks $\{A_j^{(i)}\}$ on $e_l^{(i)}$ can be formed from individual syndrome bits, which are calculated by re-encoding the (hard decision) received information symbols and adding them to the (hard decision) received parity symbols. 

For example, assuming a binary-input additive white Gaussian noise (AWGN) channel with inputs $x_l^{(i)}$ and soft outputs $y_l^{(i)}$, $i=0,1,\ldots,k$, the syndrome bit at time unit $l=0,1,\ldots$ is\vspace{-1mm}
\begin{equation}\label{eqChecksums}
s_l = \sum_{i=0}^{k-1}\sum_{b=0}^{m}\hat{y}_{l-b}^{(i)}g_{i,b}+\hat{y}_l^{(k)}= \sum_{i=0}^{k-1}\sum_{b=0}^{m}e_{l-b}^{(i)}g_{i,b}+e_l^{(k)},\vspace{-1mm}
\end{equation}
where $\hat{y}_l^{(i)}$ is the hard decision corresponding to soft output ${y}_l^{(i)}$, $e_{l}^{(i)}= \hat{y}_l^{(i)}\oplus x_l^{(i)}$ is the error symbol at time unit $l$, and $\mathbf{g}_i = (g_{i,0},g_{i,1},\ldots,g_{i,m})$ is the $i$-th code generator sequence.

In APP threshold decoding \cite{masseyAdvancesThresholdDecoding1968}, we define the \emph{reliability factor} as\vspace{-2mm}
\begin{equation}\label{eqLLR}
L(e_l^{(i)}|y_l^{(i)}) = \ln\left({\frac{\mathbb{P}(e_l^{(i)}=0|y_l^{(i)})}{\mathbb{P}(e_l^{(i)}=1|y_l^{(i)})}}\right) = 4\frac{E_s}{N_0}\left|y_l^{(i)}\right| + L(e_l^{(i)}),
\end{equation}
where $|y_l^{(i)}|$ is the channel output magnitude, $4E_s/N_0$ is  a scaled version of the channel \emph{signal-to-noise ratio} (SNR), and $L(e_l^{(i)})$ is any available \textit{a priori} information on $e_l^{(i)}$ \cite{riedelIterativeTurboDecoding1995}. Then the  $j$-th check in an orthogonal set $\{A_j^{(i)}\}$ on $e_l^{(i)}$ is weighted by the factor\vspace{-1mm}
\begin{equation}\label{eqWeights}
w_{j}^{(i)} = \sum_{\substack{\boxplus \\\alpha = 0}}^k \sum_{\substack{\boxplus \\ s \in S_j^{(i,\alpha)}}}
 L(e_s^{(\alpha)}|y_s^{(\alpha)}),\vspace{-2mm}
\end{equation}
where $\boxplus$ represents the ``box-plus" operation and $S_j^{(i,\alpha)}$ is the set of time indices involved in the $j$-th orthogonal check on the $i$-th information bit (excluding the current bit $e_l^{(i)}$), see \cite{riedelIterativeTurboDecoding1995}. 

 The APP threshold decoding rule for an orthogonal set $\{A_j^{(i)}\}$ is then given by: choose $\hat{e}_l^{(i)} = 1$ iff\vspace{-1mm}
\begin{equation}\label{eqTHRESHRULE}
L(e_l^{(i)}|\{A_j^{(i)}\}, y_l^{(i)}) = \sum_{j = 1}^{J}\left(1 -2A_j^{(i)}\right)w_{j}^{(i)}+ L(e_l^{(i)}|y_l^{(i)}) < 0,
\end{equation}

To determine the decoding decisions at time unit $l$, we add $\hat{e}_l^{(i)}$ to $\hat{y}_l^{(i)}$ to form the information symbol estimates $\hat{u}_l^{(i)}$, $i \in \{0,1, \ldots, k-1\}$.  We also add $\hat{e}_l^{(i)}$ to each syndrome equation it affects in the set of orthogonal checks $A_j^{(i)}, j \in \{1, \ldots, J\}$. The quantity  $\sum_{j = 1}^{J} (1-2A_j^{(i)})w_{j}^{(i)}$ in \eqref{eqTHRESHRULE} represents the \textit{extrinsic information} --- the \textit{a posteriori} information pertaining to all the bits in a check sum, excluding the information carried by the current channel value at time unit $l$ \cite{riedelIterativeTurboDecoding1995}. This extrinsic information is used in the iterative decoding scheme described in Section \ref{sec:decoding}.

\section{CSOC SC-PCC Encoding}\label{sec:secSCPCC_01a}
\begin{figure}[t]
\centering
\includegraphics[width = \columnwidth]{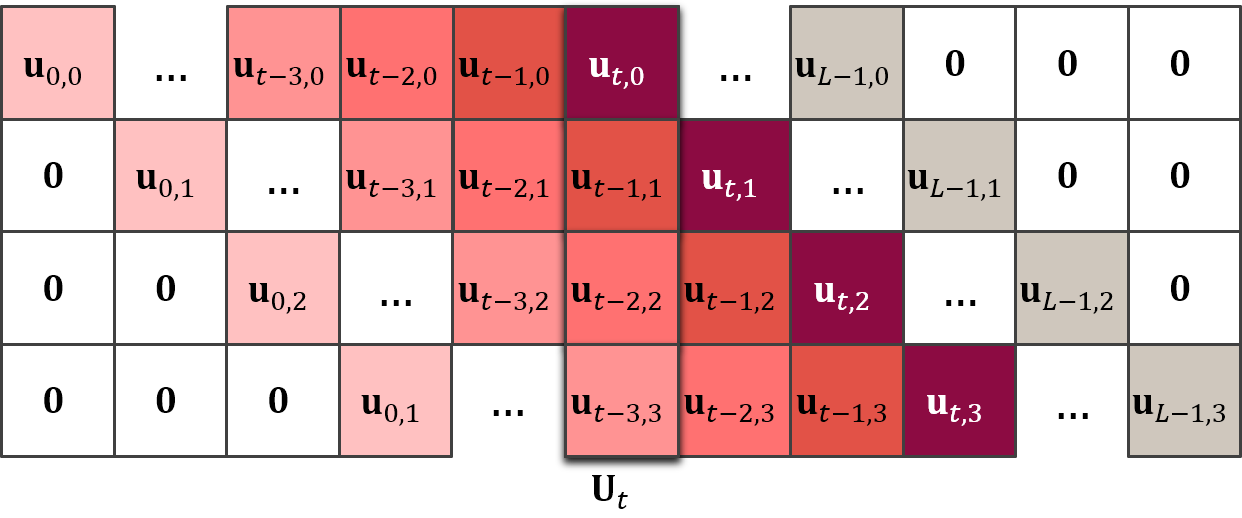}
    
\caption{Spatially coupled source matrix $\mathbf{U}$: constructing $L + m_{sc}$ coupled source blocks, each consisting of $m_{sc} + 1 = 4$ sub-blocks.}\label{SCPCC_encoding}
\end{figure}

In the proposed design, spatial coupling is accomplished by the introduction of dependencies between adjacent source information blocks prior to encoding by the component convolutional codes  \cite{moloudiSpatiallyCoupledTurboLike2017}. First, in a frame of $L$ source information blocks of size $T$ bits to be encoded, block $\mathbf{u}_t$  at time $t$ is multiplexed into $m_{sc} + 1$ separate sub-blocks $ \mathbf{u}_{t,0} , \mathbf{u}_{t,1}, \ldots , \mathbf{u}_{t, m_{sc}}$ of size ${T}/({m_{sc} + 1})$,  where $m_{sc}$ indicates the \emph{spatial coupling memory} and we assume that $T$ is a multiple of ${m_{sc} + 1}$.\footnote{By definition, a SC-PCC with $m_{sc} = 0$ is a PCC.}  The sub-blocks are then coupled over time as shown in Fig.~\ref{SCPCC_encoding}, which illustrates the process by introducing a \textit{spatially coupled source matrix} $\mathbf{U}$, where entry $\mathbf{u}_{t,i}$ of size $1\times {T}/({m_{sc} + 1})$ occupies row $i$ and column $t+i$ of $\mathbf{U}$, $i\in\{0, 1, \ldots, m_{sc}\}$, $t \in\{ 0,1,\ldots,T\}$, and column $t$ corresponds to the \emph{coupled source block} $ \mathbf{U}_t = \left(\mathbf{u}_{t,0} , \mathbf{u}_{t-1,1}, \ldots , \mathbf{u}_{t-m_{sc}, m_{sc}}\right)$. We note that $ \mathbf{U}_t$ contains sub-blocks from the source blocks at times in the interval $\left[t-m_{sc}, t\right]$ and that the first and last $m_{sc}$ coupled source blocks contain all-zero sub-blocks, as shown in Fig.~\ref{SCPCC_encoding}.

The same process is carried out for a permuted version of source block $\mathbf{u}_t$, denoted $\tilde{\mathbf{u}}_t$,  which is demultiplexed into sub-blocks $ \tilde{\mathbf{u}}_{t,0} , \tilde{\mathbf{u}}_{t,1}, \ldots , \tilde{\mathbf{u}}_{t, m_{sc}}$ of size ${T}/({m_{sc} + 1})$ and arranged into a permuted spatially coupled source matrix $\tilde{\mathbf{U}}$ (in the same format as $\mathbf{U}$), such that a permuted coupled source block $\tilde{\mathbf{U}}_t$ is produced at each time. However, we note that $\tilde{\mathbf{U}}_t$ is, in general, not a permuted version of $\mathbf{U}_t$. For example, $\mathbf{u}_{t,0}$ and $\tilde{\mathbf{u}}_{t,0}$ both consist of ${T}/({m_{sc} + 1})$ bits from source block $\mathbf{u}_{t}$, but they are likely different sets of bits due to the interleaving before demultiplexing.

\begin{figure}[t]
\centering
\includegraphics[width = \columnwidth]{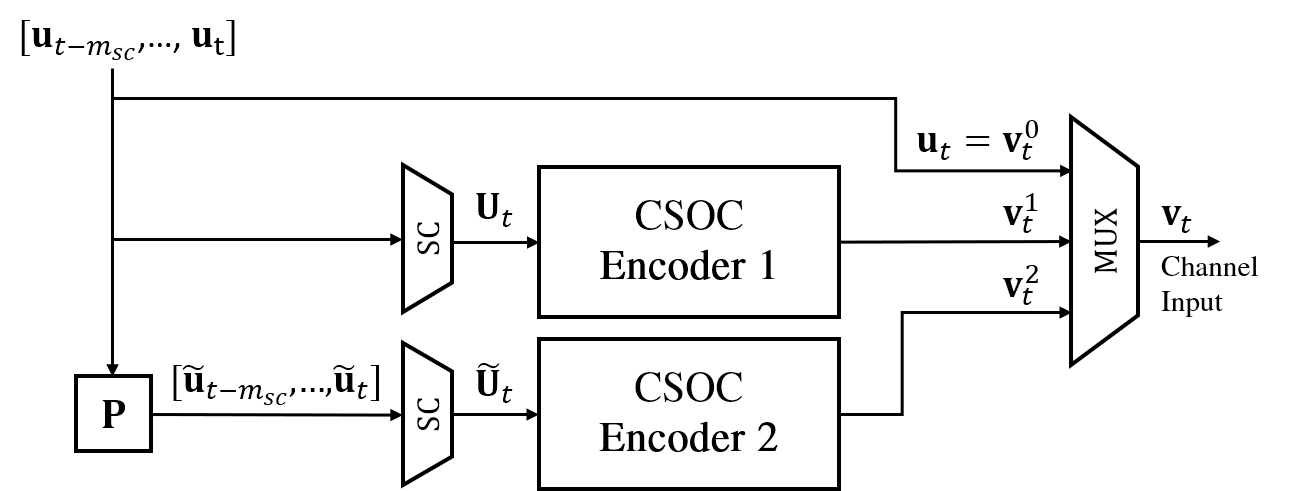}
\caption{SC-PCC encoder structure with component $(k+1,k,m)$ CSOC encoders.}\label{SCPCC_Concatcodeparallel}
\end{figure}

The blocks $\mathbf{u}_t$, $\mathbf{U}_t$, and $\tilde{\mathbf{U}}_t$ are encoded in typical turbo code fashion, as shown in Fig. \ref{SCPCC_Concatcodeparallel}, where the SC blocks correspond to the spatial coupling described above and each encoder input is terminated with $\nu \triangleq k(m +1)$ zeros to become  length $T + \nu$ input blocks to two identical feedforward CSOC component encoders.\footnote{Note that a secondary interleaving of $\mathbf{U}_t$ and $\tilde{\mathbf{U}}_t$ was performed prior to encoding in \cite{moloudiSpatiallyCoupledTurboLike2017}, but it was not found to offer additional benefit for our construction and has therefore been omitted.} Output sequences are obtained in precisely the same way as for a PCC, where the parity output blocks $\mathbf{v}^1_t$ and $\mathbf{v}^2_t$ from each encoder are concatenated with the systematic output block $ \mathbf{u}_t = \mathbf{v}^0_t$  at each time unit to form the output block $\mathbf{v}_t = ( \mathbf{v}^0_t, \mathbf{v}^1_t, \mathbf{v}^2_t)$, with an overall code rate of $R = \frac{T}{T + (2T/k) + \nu}$, approximately $\frac{k}{k+2}$ for $T$ sufficiently large. In this way, an SC-PCC can be thought of as a series of PCCs whose inputs are spread over time and whose outputs are  correlated in time over a span of $m_{sc} +1$ blocks. 

\section{Window Threshold Decoding of CSOC SC-PCCs} \label{sec:decoding}
Now let $\mathbf{r}_t=(\mathbf{r}_t^0,\mathbf{r}_t^1,\mathbf{r}_t^2)$ be the soft-valued received block at time $t$ corresponding to the transmitted block $\mathbf{v}_t = ( \mathbf{v}^0_t, \mathbf{v}^1_t, \mathbf{v}^2_t)$. Upon receiving $m_{sc}+1$ blocks, the spatial coupling is repeated by demultiplexing the received blocks $\mathbf{r}_t^0$ into $m_{sc}+1$ sub-blocks to row $i$ and column $t+i$, $i\in\{0, 1, \ldots, m_{sc}\}$ of a noisy channel information matrix $\mathbf{Y}$ (in the same way as $\mathbf{U}$), whose columns $\mathbf{Y}_t$ correspond to $\mathbf{U}_t$ and serve as the systematic inputs to the first component decoder.  This procedure is repeated for the permuted sequence, forming the permuted noisy channel information matrix $\tilde{\mathbf{Y}}$ (in the same way as $\tilde{\mathbf{U}}$) to obtain the systematic inputs $\tilde{\mathbf{Y}}_t$ to the second component decoder.  We then define $\mathbf{I}^c_t=( \mathbf{Y}_t, \mathbf{r}_t^1)$ as the information and parity inputs to decoder 1 and $\tilde{\mathbf{I}}_t^{c} =  (\tilde{\mathbf{Y}}_t,  \mathbf{r}_t^2)$ as the information and parity inputs to decoder 2.  Note that recreating blockwise spatial coupling introduces a minimum decoding latency of $m_{sc}+1$ blocks.

\begin{figure}[t]
\centering
\includegraphics[width = 3.4in]{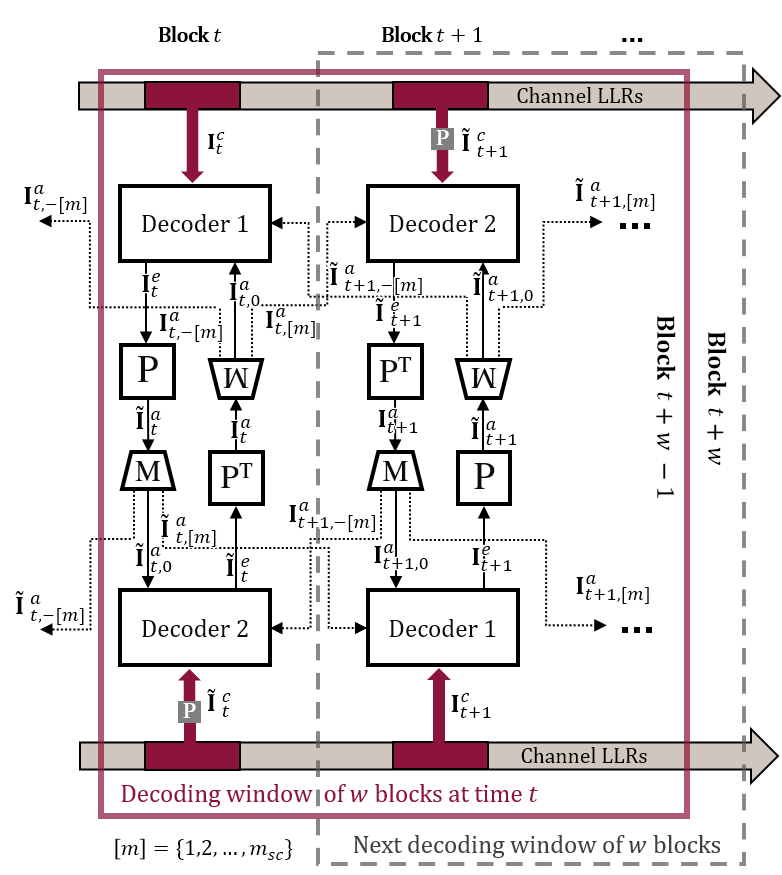}\vspace{-1mm}
\caption{Illustration of sliding window decoding, where the window covers $w$ blocks.}\label{SCPCC_decoding}\vspace{-2mm}
\end{figure}

As shown in Fig.~\ref{SCPCC_decoding}, decoding takes place within a window of size $w\geq m_{sc}+1$ blocks, with the block at the beginning of the window at time $t$ referred to as the \emph{target block}. 
At each time unit, the component decoders perform threshold decoding and exchange the extrinsic information $\mathbf{I}^e_t$ and $\tilde{\mathbf{I}}^e_t$ pertaining to the current block being decoded at time $t$, where $\mathbf{I}^e_t$ and $\tilde{\mathbf{I}}^e_t$ are produced by the component threshold decoders as the summation term of \eqref{eqTHRESHRULE} and passed to their complementary decoder as \emph{a priori} information. This turbo-decoding cycle constitutes what will be referred to as a \textit{vertical decoding iteration}. 
Starting with the target block, $I_V$ vertical decoding iterations are performed. Since spatial coupling introduces dependencies between blocks, where each decoded block depends on information from blocks at times $t-m_{sc}$ to $t+m_{sc}$, it is necessary for decoders at different time indices to exchange information. 
Therefore, the extrinsic information $\mathbf{I}^e_t$ and $\tilde{\mathbf{I}}^e_t$, produced by each decoder must be deinterleaved and demultiplexed prior to being sent to the relevant positions in the complementary decoder. 
We represent the appropriate demultiplexed \emph{a priori} information using additional subscripts as illustrated in  Fig.~\ref{SCPCC_decoding}, where Decoder 2 demultiplexes and sends $\mathbf{I}^a_{t, 0}$ to Decoder 1 at time $t$, $\mathbf{I}^a_{t, +[m]}$ is sent to the decoders to the right, and $\mathbf{I}^a_{t, -[m]}$ is sent to the decoders to the left, where $[m]=(1,\ldots,m_{sc})$ and the corresponding tilde notation applies to the demultiplexed  \emph{a priori} information sent from Decoder 1.

At the completion of $I_V$ vertical iterations, decoding continues with block $t+1$, where we see that the decoders  take advantage of the partial \emph{a priori} information obtained from the extrinsic information calculated by the complementary decoder at  time $t$.  We continue this process block-by-block until the end of the window at time $t+w-1$ is reached. 
The process then continues in the reverse direction, performing $I_V$ vertical iterations on each block from time $t+w-1$ back to the target block at time $t$, thereby completing one \emph{horizontal iteration}.  Once we have completed $I_H$ horizontal iterations, hard-decisions are made on the target symbols according to the threshold decoding rule from \eqref{eqTHRESHRULE}, and these symbols are shifted out of the decoder. 

The decoding window is now advanced one position to the right such that the target block is $t+1$ (see Fig.~\ref{SCPCC_decoding}). The process then repeats from the beginning, performing a total of $I_w = 2wI_V I_H$ vertical iterations to decode each set of target symbols until all symbols have been decoded. 

\section{Latency and Complexity Considerations}\label{sec:disc}
In this section, we  quantify the latency and implementation complexity of windowed iterative threshold decoding of SC-PCCs in comparison with iterative threshold decoding of PCCs, where we note that these suboptimal algorithms have significantly lower complexity than their optimal counterpart, i.e., iterative BCJR decoding. This serves the overall goal of this work, which is to investigate the effect of spatial coupling on low complexity, high speed, iterative decoding schemes that can close the performance gap between optimal and suboptimal decoding of PCCs (demonstrated in Sec. \ref{sec:numerical}). \vspace{-5mm}

\subsection{Latency}
Increasing the spatial coupling memory $m_{sc}$ increases both the encoding and decoding latency because $m_{sc}+1$ source blocks are needed to construct the coupled blocks $\mathbf{U}_t$ and $\mathbf{Y}_t$.  The total latency therefore depends on the chosen values of  spatial coupling memory $m_{sc}$, window size $w$, and information block size $T$, resulting in a minimum latency of $(m_{sc}+1)T$ symbols.  
The required decoding latency is ${\Delta}_d = wT = 2T(m_{sc}+1)$ symbols, with the minimum achievable latency at $m_{sc} = 1$ therefore being ${\Delta}_d = 4T$ symbols compared to $T$ symbols for an uncoupled PCC.

\subsection{Complexity} 
The complexity of the iterative APP threshold decoding algorithm was previously quantified within the context of PCCs with CSOC component codes in \cite{riedelIterativeTurboDecoding1995}. The complexity of the algorithms presented here are analyzed similarly. We will assume throughout that each encoder/decoder contains two component encoders/decoders and we disregard all elements except buffers, interleavers, and encoders/decoders, since these three components consume the vast majority of memory and computational cycles. 

\subsubsection{Memory}
The memory needed for an SC-PCC encoder is the same as that needed for a PCC encoder, but spatial coupling prior to encoding requires at least one buffer of size $T (m_{sc}+1)$ to construct the matrix $\mathbf{U}$. 
Furthermore, each encoder requires a total of $\nu$ memory elements, for a total of $2\nu$.  This gives us a required SC-PCC encoder memory of $M^\textrm{e}_\textrm{SC-PCC} \approx T (m_{sc}+1) +  2\nu$, whereas a PCC encoder requires only $T + 2\nu$ memory elements.

For an SC-PCC decoder, we need $T w$ memory elements to store extrinsic information for transfer between decoders as well as $T$ memory elements for the interleaver.\footnote{We assume deinterleaving utilizes the same memory block as interleaving read in the reverse order, and thus we only count an interleaver/deinterleaver pair as one unit.  We also assume component encoders/decoders can share the memory used for information transfer.} Each component decoder requires a register of size $2\nu$ for forming new syndromes and storing previously calculated syndromes, as well as $2\nu$ registers for storing soft channel values. This gives us a total required memory for an SC-PCC decoder of $M^\textrm{d}_\textrm{SC-PCC}  \approx Tw + T + 2(2\nu + 2\nu) = T(w+1) + 8\nu$ compared to  $T + 8\nu$ for a PCC decoder.
\subsubsection{Computation}
To simplify the discussion of the number of computations required by the component encoders and decoders, it is helpful to determine an expression for the approximate number of non-zero terms $\tilde{N_{\varnothing}}$ in the generator sequences of CSOC component codes, which is directly related to the number of computations performed when encoding and decoding.  We have found the relationship between constraint length and non-zero generator terms to be roughly linear for CSOCs, obeying the empirical formula $\bar{N_{\varnothing}} \approx \frac{\nu}{1.5k(k-1)}$.  We therefore define the scaling constant $\gamma \triangleq \left\lfloor{{1}/({1.5k(k-1)})}\right\rfloor$, which we will use hereafter in our determination of computational complexity. We only consider additions, multiplications, and box-plus ($\boxplus$) calculations, as these represent the most difficult operations.\footnote{Note, the approximate box-plus
 ($\boxplus$) operation requires only comparisons and signum function operations \cite{riedelIterativeTurboDecoding1995}.} Furthermore, we express all values in terms of computations required per $T$ bits decoded.
 
For a component decoder, each  received channel soft-value requires one multiplication to scale by $4E_s/N_0$ and one addition to add \textit{a priori} information as in \eqref{eqLLR}. Since syndromes can be formed by re-encoding the received (hard decision) information symbols and adding  to the received (hard decision) parity symbols, as in \eqref{eqChecksums}, each new syndrome  requires approximately $\nu \gamma$ additions and the same number of box-plus comparisons to calculate the weighting factors in \eqref{eqWeights}. Finally, the threshold decoding rule \eqref{eqTHRESHRULE} requires summing $kJ \approx \nu \gamma$ of these weighting factors, while the application of the decoding rule requires $k$ additions.  Breaking these down by operation, we have a total of $C^\textrm{d}_\textrm{mul} \approx T(k+1)$ multiplications, 
$C^\textrm{d}_\textrm{add} \approx T(k+1) + T(\nu \gamma) + T(\nu \gamma) = T(k + 2\nu \gamma + 1)$ additions, 
and
$C^\textrm{d}_{\boxplus} \approx T(\nu \gamma)$ 
box-plus operations, contributing to a total computational count of 
$C^\textrm{d} \approx T(2k + 3\nu\gamma + 2)$
operations required by each component decoder for every $T$ bit block.

For the SC-PCC decoder, each component decoder operates once per vertical iteration and $I_V$ vertical iterations are performed $2I_H$ times in each window position.  Thus, we multiply each component decoder computational operation count by $I_w$ to derive the total operation count 
$C^\textrm{d}_\textrm{SC-PCC} \approx 2wI_V I_H   T(2k + 3\nu \gamma + 2)$
required by the overall SC-PCC iterative decoding process. 

\subsubsection{Parallelism}
During APP threshold decoding, $k$ estimates corresponding to the $k$ information symbols are produced at each time unit.  Since these estimates are not fed back to correct the syndrome registers until all $k$ estimates are made, the decoding of each of these $k$ symbols can be performed in parallel.  This implies that the amount of allowable parallelism grows with the code rate, $R = \frac{k}{k+1}$, of the underlying component codes, making this a particularly attractive feature for high-throughput, low-overhead applications. 

\section{Numerical results}\label{sec:numerical}\vspace{-1mm}
In this section, we report binary-input AWGN channel with BPSK modulation computer simulation results obtained for the proposed iterative threshold decoding of SC-PCCs and compare its performance to uncoupled PCCs.\vspace{-1mm}

\subsection{Coupling Gain}\label{sec:coupling}
\begin{figure}[t]
\centering
\includegraphics[width = 3.2in]{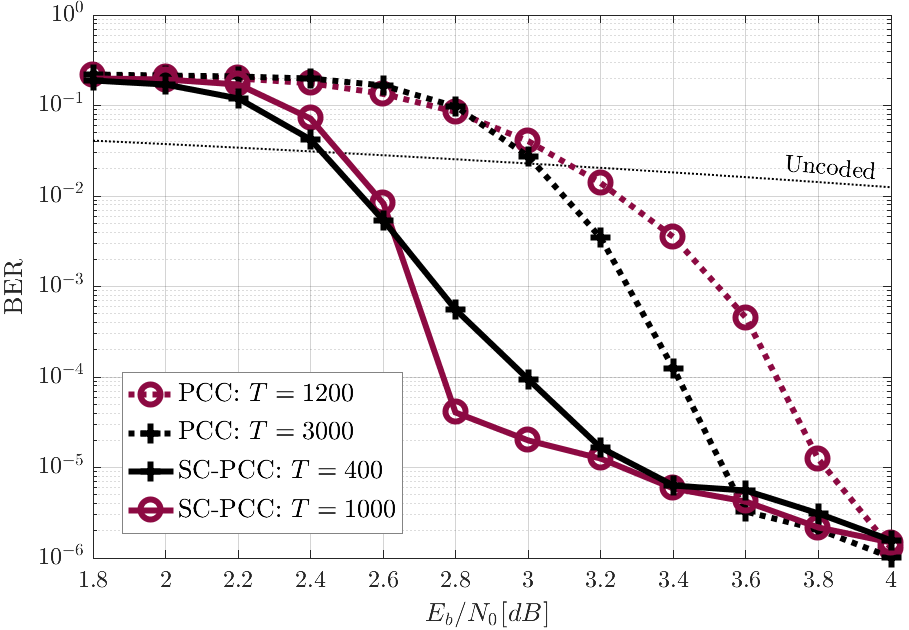}
\caption{Iterative threshold decoding of $R=1/2$ SC-PCCs and PCCs with varying block sizes $T$.}\label{SCPCC_memory}
\end{figure}

\vspace{-1mm}
Because a PCC is simply an SC-PCC with $m_{sc}=0$, we can directly compare iterative threshold decoding of a PCC and an SC-PCC with identical component codes in order to observe the impact of spatial coupling.  
Fig.~\ref{SCPCC_memory} shows simulation results for an $R = 1/2$ PCC comprised of two identical $R_c=2/3$ component codes \cite{riedelIterativeTurboDecoding1995} which have $J=4$, $m=13$, generator sequences $\textbf{g}_0 = (1001100000001)$ and $\textbf{g}_1 = (10100001000001)$, and block sizes of $T=1200$ and $3000$, along with SC-PCCs employing the same component codes with spatial coupling memory $m_{sc} = 1$, block sizes $T=400$ and $1200$, and decoding window size $w = 3$. The number of received symbols required to begin decoding for the $T=1200$ PCC and $T=400$ SC-PCC are therefore the same,\textit{ i.e.}, for the SC-PCC $\Delta_d = wT = 3\cdot 400 = 1200$. This is also true of the $T=3000$ PCC and $T=1000$ SC-PCC. The PCC decoders were allowed $I_V=24$ iterations, beyond which there is negligible performance improvement, and the SC-PCC decoders were allowed $I_V=1$ and $I_H=4$  iterations, for a matching total of $I_w = 2wI_V I_H = 24$ vertical iterations, making the PCC and SC-PCC pairs equivalent in terms of both latency and complexity. 

We observe that, even though the error floor is not improved, each SC-PCC achieves a significant performance improvement over the equal latency counterpart PCC in the waterfall. In going from just $m_{sc} = 0$ to $m_{sc} = 1$, the $T=400$ SC-PCC curve displays a gain over the PCC with $T=1200$ of approximately $0.7$dB at a bit error rate (BER) of $10^{-3}$, while the $T=1000$ SC-PCC curve shows a gain over the PCC with $T=3000$ of approximately $0.8$dB at a BER of $10^{-4}$. Because small values of $m_{sc}$ produce substantial gains over PCCs, SC-PCCs present an attractive choice compared to uncoupled alternatives. Larger memory component codes and/or larger spatial coupling memory can be selected than those chosen for this baseline comparison to \cite{riedelIterativeTurboDecoding1995} in order to further improve the performance (see Sec.~\ref{HCR}). 

\subsection{Decoding Window Size}
\begin{figure}[t]
\centering
\includegraphics[width = 3.1in]{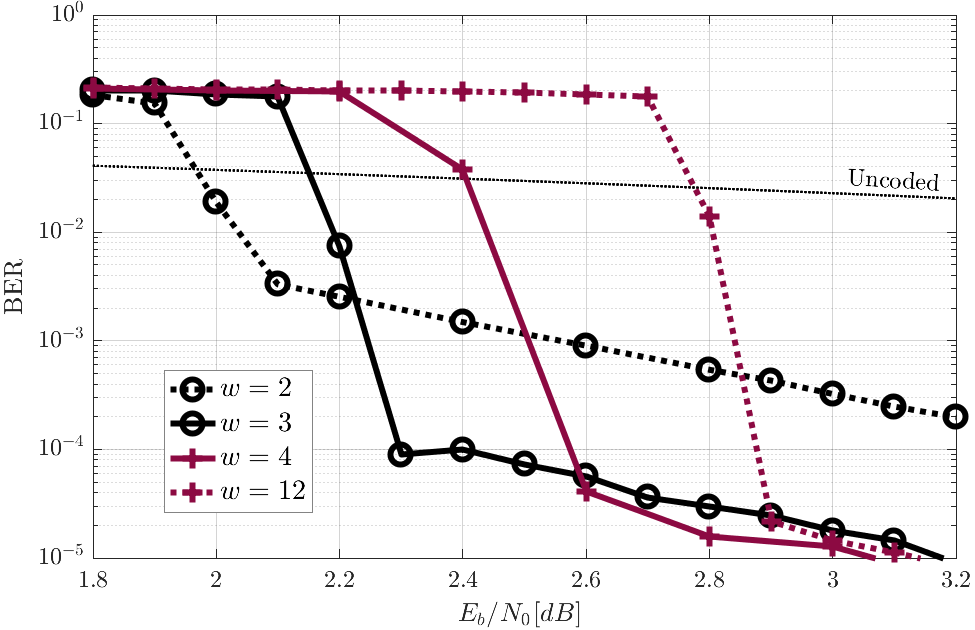}\vspace{-1mm}
\caption{SC-PCC iterative threshold decoding with increasing window size $w$.}\label{SCPCC_window}\vspace{-1mm}
\end{figure}

Fig.~\ref{SCPCC_window} illustrates the effect of varying the window size $w$ for the $R=1/2$ SC-PCC described above with $m_{sc}=1$ and $T=9990$. When the window is too small, the performance is degraded, as can be seen in Fig.~\ref{SCPCC_window} for $w=2$.  Increasing the window size further improves the performance; however, window sizes much larger than the spatial coupling memory, \textit{i.e.}, $w \gg m_{sc}$, offer no additional benefit (\emph{e.g.}, $w =12$).\footnote{The observed waterfall performance degradation for large $w$ is due in part to the schedule (vertical and horizontal updates) and in part to the  unreliable and large valued extrinsic information generated by the threshold decoding of blocks at the far right end of the decoding window when the SNR is not sufficiently large. This effect can be mitigated by appropriate scaling of the extrinsic information produced by the decoder.} Based on these observations, we see that the decoding window size has a significant impact on decoder performance, and we have determined empirically that $w_d \approx 2(m_{sc}+1)$ serves as a good default value for this parameter to trade off waterfall and error floor performance.  \vspace{-1mm}

\subsection{Higher Code Rates}\label{HCR}
\begin{figure}[t]
\centering
\includegraphics[width =3.1in]{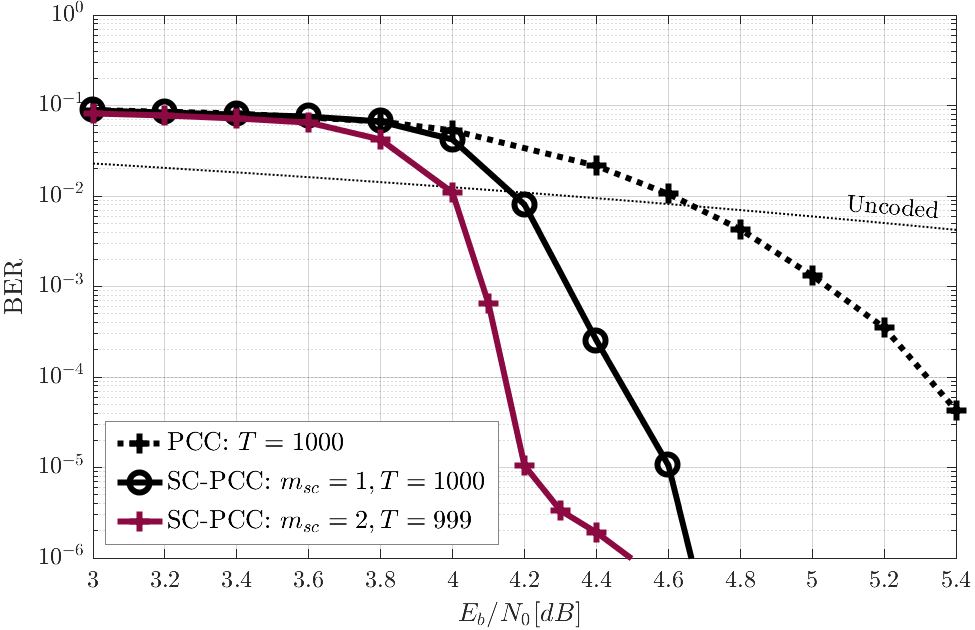}\vspace{-1mm}
\caption{Iterative threshold decoding of rate $R=4/5$ SC-PCC and PCC.}\vspace{-1mm}\label{fig:highrate}
\end{figure}

An advantage of the proposed iterative threshold decoding of SC-PCCs is that high rate and high memory CSOC component codes can be used with only a modest increase in complexity (as quantified in Section \ref{sec:disc}). Fig. \ref{fig:highrate} shows the performance of an $R=4/5$ PCC constructed from two identical $R_c=8/9$, $J=4$, $m=136$ component codes \cite{wuNewConvolutionalCodes1975} along with the corresponding SC-PCC code with $m_{sc} = 1$ and $2$ and $w=w_d=4$, all with $T=1000$.\footnote{Terminating every block introduces a substantial rate loss when the ratio $T/m$ is small (in this example, the actual rate is $R=0.656$).  This block termination can be omitted for the SC-PCC since the sliding window decoder can share information about symbols at the beginning and end of each block with component decoders at adjacent times. 
} Here, we set $I_V = 16$ for the PCC decoder and $I_V = 4$ and $I_H = 2$ for the SC-PCC decoders. Similar to the $R=1/2$ examples above, we observe significant improvement from spatial coupling, achieving a gain of approximately $1$dB with $m_{sc} = 1$ and $1.4$dB with $m_{sc} = 2$ at a BER of $10^{-4}$.\vspace{-1mm}

\section{Concluding Remarks and Future Directions}
In this paper, we have considered spatial coupling of parallel concatenated, turbo-like codes  and have developed a low complexity, iterative decoding algorithm based on Massey's APP threshold decoder.  Though such SC-PCC constructions do not offer a lower error floor than their uncoupled PCC counterparts, significant gains in the waterfall region were demonstrated and the simplicity of the decoder lends itself to large memory and high rate component codes  for high-throughput, low-overhead applications.

The techniques explored in this paper can be extended by replacing the component CSOCs with stronger doubly-self orthogonal convolutional codes (CSO2Cs) codes, whose structure allows for repeated application of the threshold decoding rule to the same code block \cite{roySimplifiedHighRatePunctured2007}. It is also possible to extend the iterative threshold decoding framework  to other classes of SC-TCs, such as laminated turbo codes \cite{hzc08}, spatially coupled, serially concatenated convolutional codes (SC-SCCs) \cite{moloudiSpatiallyCoupledTurboLike2017}, braided convolutional codes (BCCs) \cite{zhuBraidedConvolutionalCodes2017a}, partially information coupled (PIC) turbo codes \cite{yangPartiallyInformationCoupledTurbo2018}, and staircase codes \cite{zhangFeedForwardStaircaseCodes2017}.  For the classes of SC-TCs that incorporate parity feedback in the encoder, the underlying APP threshold decoding algorithm must be adapted to provide parity extrinsic information and the decoder structure must be modified accordingly.\vspace{-1mm}

\vspace{-0.4mm}

%%%%%%%%%%%%%%%%%%%%%%%%%%%%%%%%%%%%
%\section*{Acknowledgment}\label{sec:secAck}
%%%%%%%%%%%%%%%%%%%%%%%%%%%%%%%%%%%%

\bibliographystyle{IEEEtran}

\end{document}